\documentclass[aps,prd,twocolumn,showpacs,nofootinbib]{revtex4}

\usepackage{graphicx}

\begin{document}

\title{Reanalysis of the Higher Order Perturbative QCD corrections to Hadronic $Z$ Decays using the Principle of Maximum Conformality}
\author{Sheng-Quan Wang}
\email{sqwang@cqu.edu.cn}

\author{Xing-Gang Wu}
\email{wuxg@cqu.edu.cn}
\address{Department of Physics, Chongqing University, Chongqing 401331, P.R. China}

\author{Stanley J. Brodsky}
\email{sjbth@slac.stanford.edu}
\affiliation{SLAC National Accelerator Laboratory, Stanford University, Stanford, CA 94309, USA}

\date{\today}

\begin{abstract}
A complete calculation of the ${\cal O}(\alpha_s^4)$  perturbative QCD corrections to the hadronic decay width of the $Z$-boson has recently been performed by Baikov et al.~\cite{rns42}. In their analysis, Baikov et al. relied on the conventional practice of simply guessing the renormalization scale and taking an arbitrary range to estimate the pQCD uncertainties.  This procedure inevitably introduces an arbitrary, scheme-dependent theoretical systematic error in the predictions. In this paper, we show that the renormalization scale uncertainties for hadronic $Z$ decays can be greatly reduced by applying the principle of maximum conformality (PMC), a rigorous extension of the BLM method. The PMC prediction is independent of the choice of renormalization scheme; i.e., it respects  renormalization group invariance, and thus it provides an optimal and theoretically rigorous method for setting the renormalization scale. We show that the convergence of the pQCD prediction for the $Z$ hadronic width is greatly improved using the PMC since the divergent renormalon series does not appear.  The magnitude of the high-order corrections quickly approach a steady point. The PMC predictions also have the property that any residual dependence on the choice of initial scale is highly suppressed, even for low-order predictions. Thus, one obtains optimal fixed-order predictions for the $Z$-boson hadronic decay rates thus enabling high precision tests of the Standard Model.
\end{abstract}


\maketitle

A complete calculation of the ${\mathcal{O}}(\alpha_s^4)$ QCD corrections to the hadronic decay width of the $Z$-boson has recently been reported in Ref.\cite{rns42}. The computation of the new high-order terms, especially the four-loop terms for the dominant non-singlet part can lead to a significant improvement in the precision of theoretical prediction for the $Z$-boson decay width and thus a new level of precision for the determination of $\alpha_s(M_Z)$. However, the pQCD analysis in Ref.\cite{rns42} uses the conventional practice of simple guessing the renormalization scale and taking an arbitrary range to estimate the pQCD uncertainties. This procedure inevitably introduces an arbitrary theoretical systematic error in the predictions. For example, the estimated uncertainties in the components to the decay width are $\delta\Gamma_{\rm NS}$=0.101 MeV, $\delta\Gamma^V_{S}$=0.0027 MeV, and $\delta\Gamma^{A}_{S}$=0.042 MeV if one varies the renormalization scale in the range $\mu_r\in[M_Z/3, 3M_Z]$~\cite{rns42}. An additional theoretical problem is that two types of logarithmic terms appear related to the renormalization of the axial singlet contribution; i.e. $\ln(\mu_r/M_{Z})$ and $\ln(\mu_r/M_{t})$. Thus the usual assumption that the renormalization scale only depends on $M_Z$, but not the top quark mass, does not have a clear justification.

A valid prediction for a physical observable should not depend on theoretical conventions such as the choice of the renormalization scheme or the initial choice of the renormalization scale. This important principle, called ``renormalization group invariance (RGI)" can be satisfied at every finite order of perturbation theory using the ``Principle of Maximum conformality (PMC)~\cite{pmc1,pmc2,pmc3,pmc4,pmc5}". The PMC provides a rigorous extension of the BLM method~\cite{blm}. When one applies the PMC, the scales of the QCD coupling are shifted at each order in perturbation theory such that no contributions proportional to the QCD $\beta$ function remain. The resulting pQCD series after applying the PMC is thus identical to the series of the corresponding conformal theory. The resulting series is thus scheme independent, free of divergent $n! \alpha_s^n \beta^n$ renormalon terms. One also finds that the prediction is independent of the choice of the initial renormalization scale to very high accuracy. All properties of the renormalization group are satisfied by the PMC. The same principle underlies the Gell Mann-Low scale setting procedure used for precision tests of QED.

The PMC provides a systematic and unambiguous procedure to set the renormalization scale for any QCD process~\cite{pmc1,pmc2,pmc3,pmc4,pmc5}, thus greatly improving the precision of tests of the Standard Model. In the present paper, we shall show that the renormalization scale uncertainties given in Ref.\cite{rns42} for hadronic $Z$ decays are greatly reduced by applying the PMC.

The decay rate of the $Z$-boson into hadrons can be expressed as,
\begin{displaymath}
\Gamma_Z = {\cal C}\bigg[\sum\limits_{f}v^2_f r^V_{\rm NS}+ \bigg(\sum\limits_{f}v_f\bigg)^2 r^V_S +\sum\limits_{f} a^2_f r^A_{\rm NS}+r^A_S\bigg],
\end{displaymath}
where the factor ${\cal C}=3 \frac{G_{F}M^3_Z} {24\pi\sqrt{2}}$, $v_f\equiv 2I_f-4q_fs^2_W$, $a_f\equiv2I_f$, $q_f$ is the $f$-quark electric charge, $s_W$ is the effective weak mixing angle, and $I_f$ is the third component of weak isospin of the left-handed component of $f$. $r^V_{\rm NS}=r^A_{\rm NS}\equiv r_{\rm NS}$, $r^V_S$ and $r^A_S$ stand for the non-singlet, vector-singlet and axial-singlet part, respectively. These contributions can be further expressed as
\begin{displaymath}
r_{\rm NS} = 1+\sum^n_{i=1}C^{\rm NS}_ia^i_s, r^V_S =\sum^n_{i=3}C^{\rm VS}_{i}a^i_s, ~r^A_S=\sum^n_{i=2}C^{\rm AS}_{i}a^i_s ,
\end{displaymath}
where $a_s=\alpha_s/\pi$. The coefficients of $r_{\rm NS}$, $r^V_S$ and $r^A_S$ with their dependence on the choice of initial scale can be derived from Refs.\cite{rns42,rns41,rvs41}. Each physical contribution to the $Z$ decay has a different momentum flow, thus, the PMC scales for $r_{\rm NS}$, $r^V_S$ and $r^A_S$ should be set separately. More explicitly, up to the $\mathcal{O}(\alpha_s^4)$ level, we will have four PMC scales for $r_{\rm NS}$, i.e. $Q^{\rm NS}_{1,\cdots,4}$ for LO, NLO, N$^2$LO or N$^3$LO level, respectively; we will have two PMC scales for $r^V_S$, i.e. $Q^{\rm VS}_1$ and $Q^{\rm VS}_2$. As for $r^A_S$, we adopt previous results at the order ${\mathcal{O}}(\alpha_s^3)$~\cite{rns31,rns32} to perform our analysis\footnote{Because of a lack of information, we cannot derive the full expression for $r^A_S$ with explicit $n_f$-dependence and initial scale dependence for the newly obtained ${\mathcal{O}}(\alpha_s^4)$-terms from Ref.\cite{rns42}.}. Because $r^A_S$ provides a quite small contribution in comparison to the dominant $r_{\rm NS}$, this treatment will not affect our final conclusions. Then, we will have two PMC scales, i.e. $Q^{\rm AS}_1$ and $Q^{\rm AS}_2$, for $r^A_S$. Moreover, we do not have knowledge of the $\beta$-terms to set the PMC scales for highest order terms at ${\mathcal{O}}(\alpha_s^4)$; thus for as a convention; we will set their scales equal to the last known PMC scale; i.e. $Q^{\rm NS}_4$=$Q^{\rm NS}_3$, $Q^{\rm VS}_2$=$Q^{\rm VS}_1$ and $Q^{\rm AS}_2$=$Q^{\rm AS}_1$. This treatment will lead to some residual scale dependence, which, however, is highly suppressed.

In order to obtain numerical results, we take the top-quark pole mass $M_{t}=173.3$ GeV~\cite{toppole} and the $Z$-boson mass $M_{Z}=91.1876$ GeV~\cite{pdg}. For self-consistency, we shall use the $n$-loop $\alpha_s$-running behavior to calculate $\Gamma_Z$ up to the $n$-loop level. For this purpose, by taking $\alpha_s(M_{Z})=0.1184$~\cite{pdg}, we obtain $\Lambda^{(n_f=5)}_{\rm QCD}=0.0899$ GeV for one-loop $\alpha_s$-running, $\Lambda^{(n_f=5)}_{\rm QCD}=0.231$ GeV for the two-loop $\alpha_s$-running, and $\Lambda^{(n_f=5)}_{\rm QCD}=0.213$ GeV for the three-loop and four-loop $\alpha_s$-running, respectively.

\begin{figure}[htb]
\includegraphics[width=0.48\textwidth]{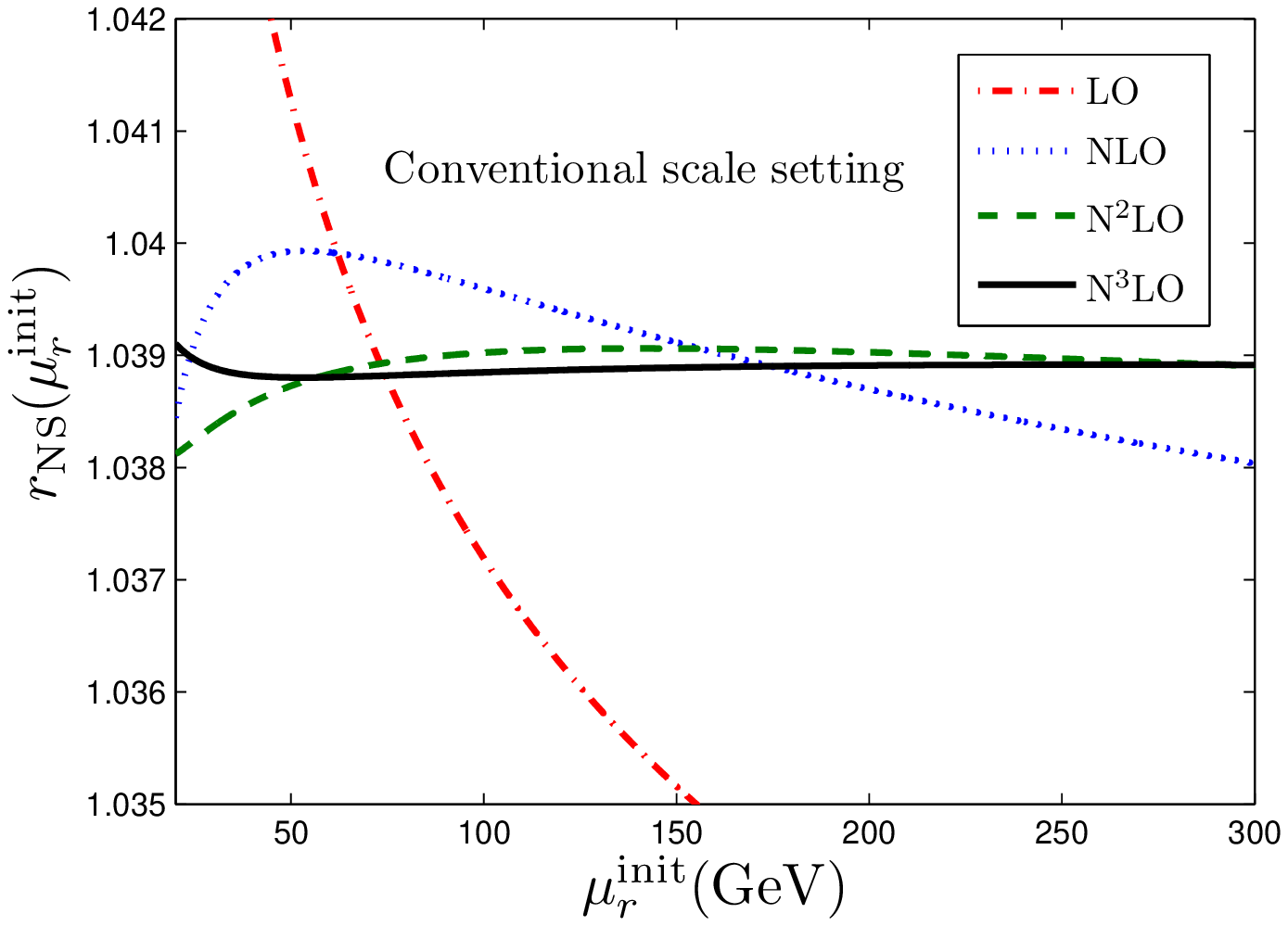}
\includegraphics[width=0.48\textwidth]{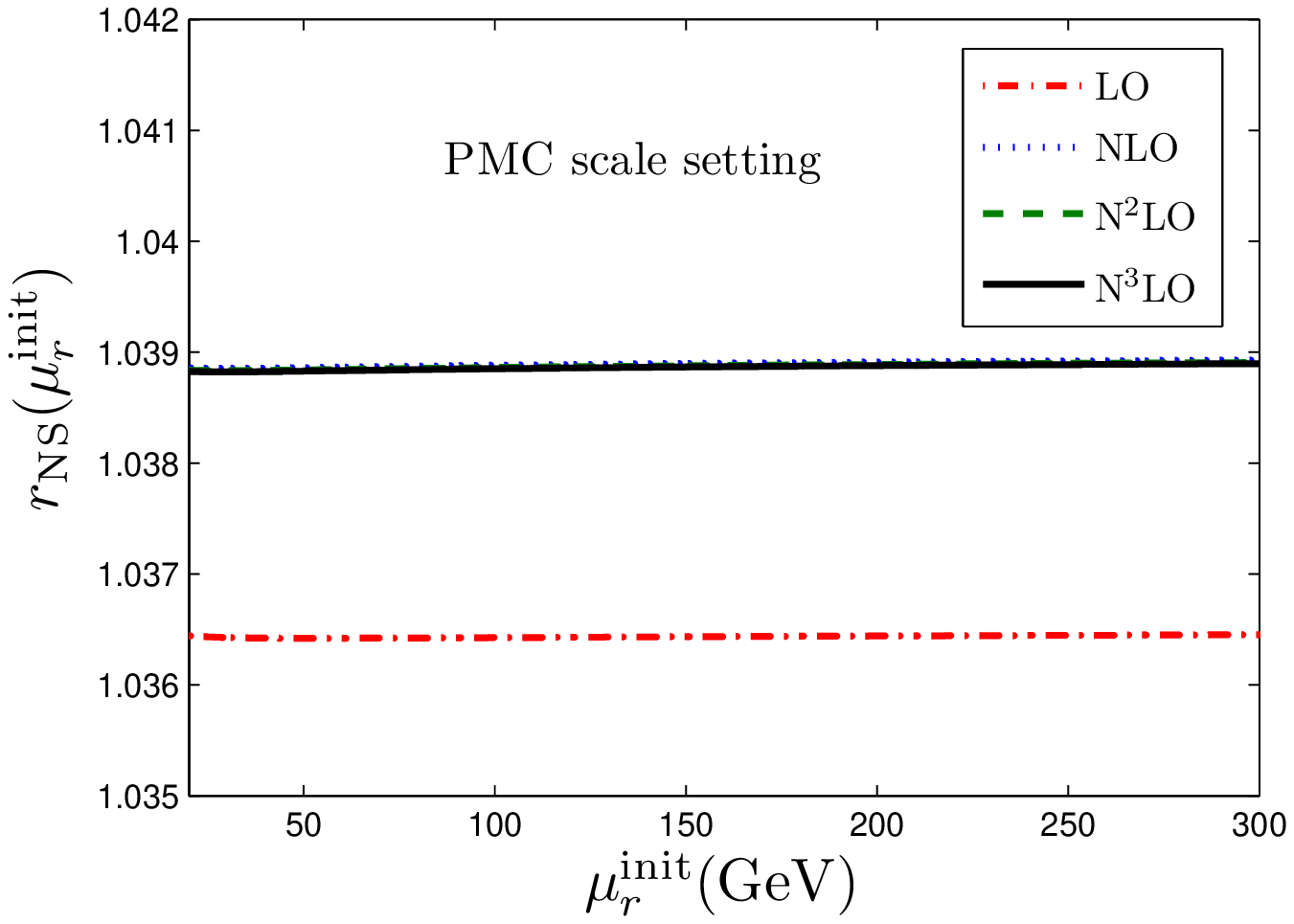}
\caption{The non-singlet contribution $r_{\rm NS}$ up to four-loop QCD corrections versus $\mu_r^{\rm init}$ before and after PMC scale setting. After PMC scale setting, the curves for NLO, N$^2$LO, and N$^3$LO are almost coincide with each other.}
\label{Rnscon_pmc}
\end{figure}

\begin{figure}[htb]
\includegraphics[width=0.48\textwidth]{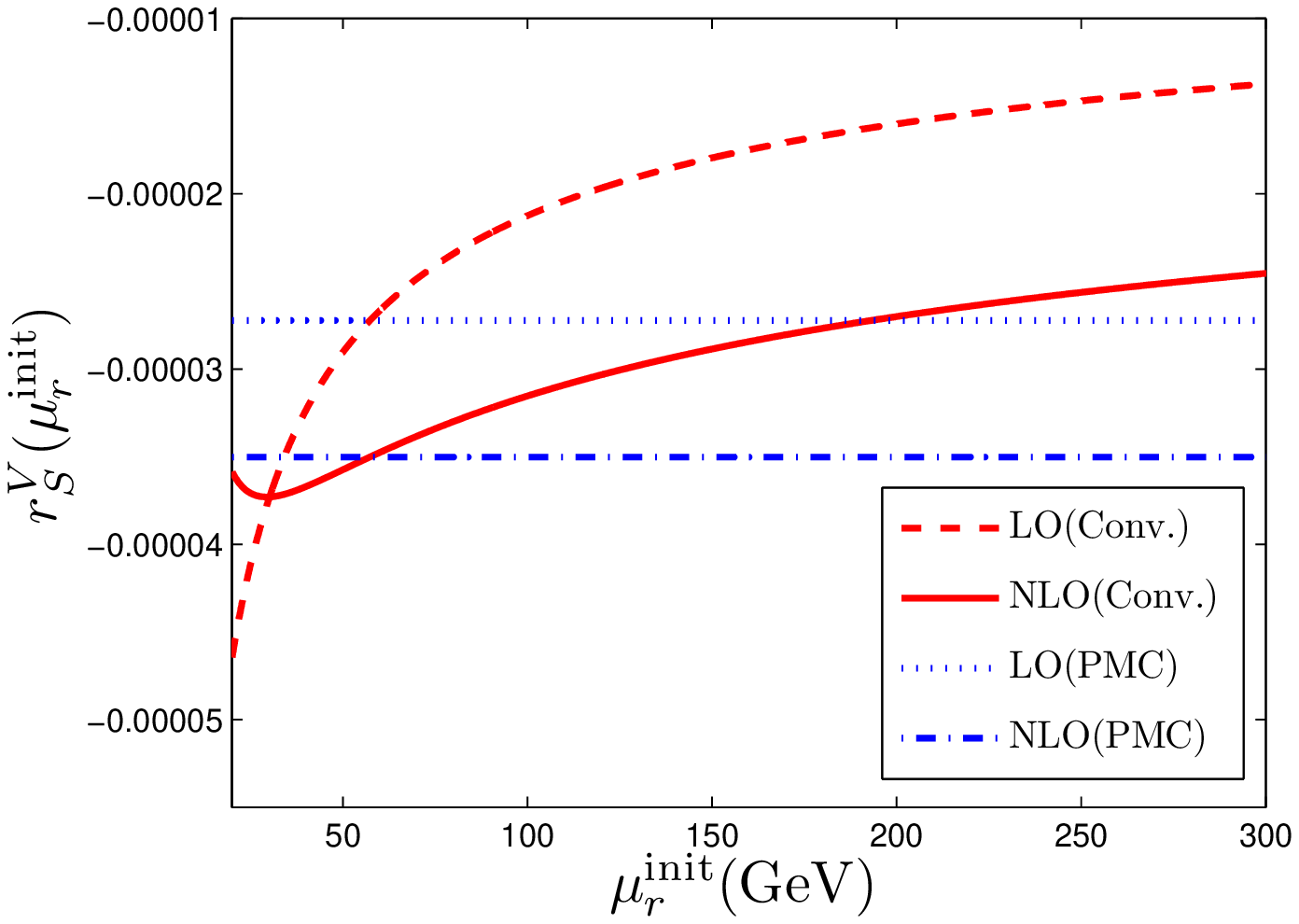}
\caption{The vector-singlet contribution $r^V_S$ up to two-loop QCD corrections versus the initial renormalization scale $\mu_r^{\rm init}$ before and after PMC scale setting.}
\label{Rvscon_pmc}
\end{figure}

\begin{figure}[htb]
\includegraphics[width=0.48\textwidth]{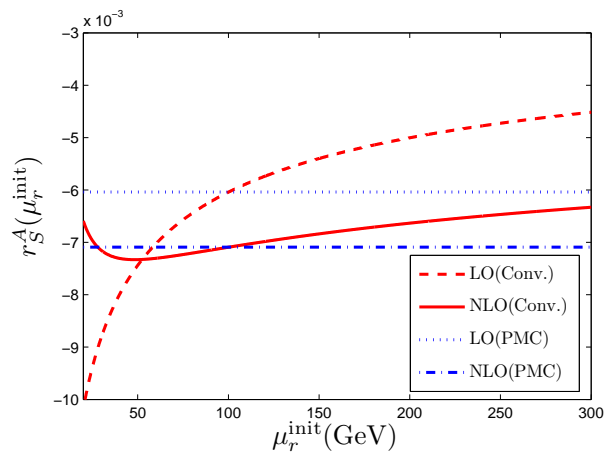}
\caption{The axial singlet contribution $r^A_S$ up to two-loop QCD corrections versus the initial renormalization scale $\mu_r^{\rm init}$ before and after PMC scale setting.}
\label{Rascon_pmc}
\end{figure}

The pattern of $\beta$ terms in a pQCD series can be systematically deduced by using the $R_\delta$ method discussed in Refs.\cite{pmc3,pmc4,pmc5}. As emphasized above, the PMC scales at each perturbative order are then determined unambiguously by absorbing all non-conformal $\beta$-terms into the running coupling; the resulting pQCD series is identical to that of the conformal theory with $\beta=0$ and is thus scheme independent. The PMC scales correctly characterize the virtuality of the propagating gluons and thus also allow one to determine the value of the effective number of flavors $n_f$. We present the (initial) scale dependence before and after PMC scale setting for $r_{\rm NS}$, $r^V_S$, and $r^A_S$ in Figs.(\ref{Rnscon_pmc},\ref{Rvscon_pmc},\ref{Rascon_pmc}).

When using conventional scale setting, i.e. $\mu_r\equiv\mu^{\rm init}_r$, the resulting low-order predictions depend strongly on $\mu^{\rm init}_r$. One does observe that as more loops are taken into consideration, one obtains a weaker scale dependence. On the other hand, after applying PMC scale setting, the pQCD predictions at each perturbative order are almost independent of the initial scale $\mu_r^{\rm init}$. This indicates that the PMC predictions have the property that any residual dependence on the choice of initial scale is highly suppressed, even for low-order predictions. For example, Fig.(\ref{Rnscon_pmc}) shows not only that the renormalization scale ambiguities are eliminated, but also that the value of $r_{\rm NS}$ rapidly approaches its steady point; i.e; the curves at NLO, N$^2$LO, and N$^3$LO almost coincide with each other after applying PMC.

A prediction for a physical observable should not depend on the initial choice of the scale. In fact, one sees that the computed PMC scales and thus the final PMC predictions for the hadronic $Z$ decay results presented here are highly independent of this choice. We find that the PMC scales for $r_{\rm NS}$ are $Q^{\rm NS}_1\simeq115.9$ GeV, $Q^{\rm NS}_2\simeq102.8$ GeV and $Q^{\rm NS}_3\simeq486.8$ GeV; the PMC scale for $r^V_{S}$ is $Q^{\rm VS}_1\simeq57.3$ GeV; the PMC scale for $r^A_{S}$ is $Q^{\rm AS}_1\simeq100.1$ GeV. We note that if one sets $\mu_r^{\rm init}\sim M_Z/2$ using conventional scale setting for the $r^V_{S}$ part, one obtains almost the same results as that of the PMC scale setting. This illustrates that the effective momentum flow for $r^V_{S}$ is $\sim M_Z/2$, rather than the conventional guess $M_Z$. We also note that both types of logarithmic-terms $\ln(\mu_r^{\rm init}/M_{Z})$ and $\ln(\mu_r^{\rm init}/M_{t})$ for $r^A_S$ are consistently incorporated by PMC scale setting. The derived value $Q^{\rm AS}_1\simeq 100$ GeV shows that the typical momentum flow for $r^A_S$ is closer to $M_Z$ than $M_t$, providing after-the-fact justification for the conventional choice of $M_Z$.

\begin{table}[htb]
\centering
\begin{tabular}{|c|c|c|c|c|c|}
\hline
~~~ ~~~ & $R^{\rm NS}_{1}$ & $R^{\rm NS}_{2}$ & $R^{\rm NS}_{3}$ & $R^{\rm NS}_{4}$  &~$\sum^{4}_{i={1}} R^{\rm NS}_{i}$~ \\
\hline
Conv. &  0.03769  & 0.00200 & -0.00069 & -0.00016 & 0.03884 \\
\hline
PMC &  0.03636 & 0.00252 & -0.00003 & -0.00001 & 0.03885\\
\hline
\end{tabular}
\caption{Perturbative contributions for the non-singlet $r_{\rm NS}$ under the conventional (Conv.) and the PMC scale settings. Here, $R^{\rm NS}_{i}=C^{\rm NS}_i a^i_s$ with $i=(1,\cdots,4)$ stand for the one-, two-, three-, and four-loop terms, respectively. $\mu_r^{\rm init}=M_Z$. } \label{table:each}
\end{table}

We emphasize that after PMC scale setting, one obtain better pQCD convergence due to the elimination of the renormalon terms. For example, the pQCD estimations at each perturbative order for $r_{\rm NS}$ are presented in Table \ref{table:each}. The fastest pQCD convergence is thus achieved by applying PMC. The pQCD correction at ${\cal O}(\alpha^4_s)$ is -0.00016 for the conventional scale setting, which leads to a shift of the central value of $\alpha_s(M_Z)$ from 0.1185 to 0.1190~\cite{rns41}. In contrast after PMC scale setting, this correction becomes $-0.00001$, which is negligible. Table \ref{table:each} also shows that the predictions for the total sum $\sum^{4}_{i=1} R^{\rm NS}_{i}$ are close in value for both PMC and conventional scale setting, although their perturbative series behave very differently.

\begin{figure}[tb]
\includegraphics[width=0.48\textwidth]{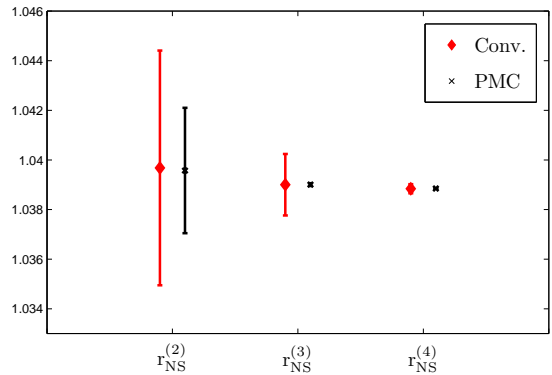}
\caption{The values of $r^{(n)}_{\rm NS}=1+\sum^n_{i=1}C^{\rm NS}_i a^i_s$ and their errors $\pm |C^{\rm NS}_n a^n_s|_{\rm MAX}$. The diamonds and the crosses are for conventional (Conv.) and PMC scale settings, respectively. The central values assume the initial scale choice  $\mu^{\rm init}_r=M_Z$. } \label{uncert}
\end{figure}

One usually estimates the unknown contributions from higher-order terms by varying $\mu_r^{\rm init}\in[M_Z/2,2M_Z]$. However, this procedure only exposes the $\beta$-dependent non-conformal terms -- not the entire perturbative series. For a conservative estimate, we take the uncertainty to be the last known perturbative order~\cite{pmc6}, i.e. at the $n$-th order the perturbative uncertainty is estimated by $\pm |C^{\rm NS}_n a^n_s|_{\rm MAX}$, where the symbol ``MAX" stands for the maximum of $|C^{\rm NS}_na^n_s|$ by varying $\mu_r^{\rm init}$ within the region of $[M_Z/2,2M_Z]$. The error bars for PMC and the conventional scale setting are displayed in Fig.(\ref{uncert}). It shows that the predicted error bars from the ``unknown" higher-order corrections quickly approach their steady points after applying PMC scale setting. Such error bars provide a consistent estimate of the ``unknown" QCD corrections under various scale settings; i.e., the exact value for the ``unknown" $r^{(n)}_{\rm NS}$  ($n=3$ and $4$) are well within the error bars predicted from the one-order lower $r^{(n-1)}_{\rm NS}$.

In Ref.\cite{rns42}, the Principle of Minimum Sensitivity (PMS)~\cite{pms} has been adopted as a guide for setting the renormalization scale. The PMS provides an intuitive way to set the renormalization scale: it breaks the standard RGI but introduces instead a local RGI to determine the optimal renormalization scale. The PMS scale is determined by requiring the slope of the approximant of an observable to vanish. Since the PMS breaks the standard RGI, the PMS does not satisfy the self-consistency conditions of the renormalization group~\cite{pmc7}. Worse, the PMS disagrees with Gell Mann-Low scale setting when applied to QED and gives unphysical results for jet production in $e^+ e^-$ annihilation~\cite{Kramer}. A detailed comparison of the PMC and PMS procedures via two physical observables $R_{e+e-}$ and $\Gamma(H\to b\bar{b})$ up to the four-loop level has been recently given in Ref.\cite{pmc6}. The PMC and PMS predictions agree with each other within quite small errors if high-enough loop corrections have been included. However, the convergence of the pQCD series at high order behaves quite differently: The PMC displays the best pQCD convergence; in contrast, the convergence of the PMS prediction is questionable, often even worse than the conventional prediction based on an arbitrary guess for the renormalization scale.

In summary, the PMC allows one to consistently and unambiguously set the renormalization scale at each order of a pQCD calculations. The nonconformal terms precisely set the renormalization scale and the conformal terms accurately display the magnitude of the pQCD correction at each order. The PMC results are independent of the choice of the renormalization scheme, as required by RGI. The PMC systematically provides the optimal scale of the QCD coupling for any process. The {\it ad hoc} error estimate usually assigned to pQCD predictions due to scale uncertainties  and scheme dependence are thus eliminated. A new, consistent method for estimating the uncertainty in a pQCD calculation based on the conformal series is also obtained. In fact, a remarkably convergent renormalon-free pQCD series emerges, as seen in Table \ref{table:each}, for the application to hadronic Z decays discussed here.

\hspace{2cm}

{\bf Acknowledgements:} This work was supported in part by Natural Science Foundation of China under Grant No.11275280, the Fundamental Research Funds for the Central Universities under Grant No.CQDXWL-2012-Z002 and the Department of Energy Contract No.DE-AC02-76SF00515. SLAC-PUB-15971.

\end{document}